\documentclass[a4paper]{article}

\usepackage[english]{babel}
\usepackage[utf8x]{inputenc}
\usepackage[T1]{fontenc}

\usepackage[a4paper,top=3cm,bottom=2cm,left=3cm,right=3cm,marginparwidth=1.75cm]{geometry}

\usepackage{amsmath}
\usepackage{algpseudocode}
\usepackage{graphicx}
\usepackage[colorinlistoftodos]{todonotes}
\usepackage[colorlinks=true, allcolors=blue]{hyperref}

\title{CoHSI II;  The average length of proteins, evolutionary pressure and eukaryotic fine structure}
\author{Les Hatton, Gregory Warr}

\begin{document}
\maketitle

\begin{abstract}
The CoHSI (Conservation of Hartley-Shannon Information) distribution is at the heart of a wide-class of discrete systems, defining (amongst other properties) the length distribution of their components.  Discrete systems such as the known proteome, computer software and texts are all known to fit this distribution accurately.  In a previous paper, we explored the properties of this distribution in detail.  Here we will use these properties to show why the average length of components in general and proteins in particular is highly conserved, howsoever measured, demonstrating this on various aggregations of proteins taken from the UniProt database.  We will go on to define departures from this equilibrium state, identifying fine structure in the average length of eukaryotic proteins that result from evolutionary processes.
\end{abstract}

\section{Statement of computational reproducibility}
To support falsifiability in the Popperian sense, this paper is accompanied by a complete computational reproducibility suite including all software source code, data references and the various glue scripts necessary to reproduce each figure, table and statistical analysis and then regress local results against a gold standard embedded within the suite to help build confidence in the theory and results we are reporting.  This follows the methods broadly described by \cite{Ince2012} and exemplified in a tutorial and case study \cite{HattonWarr2016}.  These reproducibility suites are currently available at http://leshatton.org/ until a suitable public archive appears, where they may be transferred.

\section{Introduction}
In previous papers \cite{HattonWarr2018a}, we have derived and explored the properties of a differential equation which accurately models the global length distribution of components of discrete systems. Examples would include i) the lengths of all known proteins, either for a species (the proteome) or indeed for all organisms (the pan-proteome) and ii) the lengths of functions in computer programs written in any programming language.  Throughout we will consider a discrete system as a set of components, each of which comprises a number of indivisible tokens or symbols in which order is significant, chosen from a unique alphabet of tokens.  In such a system, the $i^{th}$ component is taken to have $t_{i}$ tokens and a unique alphabet of $a_{i}$ tokens.  The total size of the system is $T = \sum_{i=1}^{M} t_{i}$ where $M$ is the number of components and is taken to be reasonably large, $M > 20$ say.  This leads to the heterogeneous model of \cite{HattonWarr2017}, which is directly applicable to the proteome.

With this nomenclature, the differential equation describing the length distribution as an implicit pdf $\sim a_{i}(t_{i})$ is

\begin{equation}
\log t_{i} + \frac{1 + 8 t_{i} + 24t_{i}^{2}}{6(t_{i} + 4 t_{i}^{2} + 8 t_{i}^{3})} = -\alpha -\beta ( \frac{d}{dt_{i}} \log N(t_{i}, a_{i}; a_{i} ) ),    \label{eq:minif}
\end{equation}

Here $N(t_{i}, a_{i}; a_{i} )$ is a function describing the Hartley-Shannon information content of the $i^{th}$ component as described in detail in \cite{HattonWarr2018a}.  The parameters $\alpha, \beta$ are Lagrangian undetermined parameters.  In other words the Statistical Mechanics methodology we used has nothing to say about their value - they are an unknown function of the discrete dataset being studied and we will discuss their relevance later.

(\ref{eq:minif}) arises naturally for systems which have the same measure of Hartley-Shannon Information for a fixed size in total number of symbols or tokens.  It is both scale-independent (it does not depend on $T$ providing T is reasonably large in the sense of Statistical Mechanics \cite{GlazerWark2001}) and token agnostic (Hartley-Shannon information specifically avoids any associated meaning to the tokens \cite{Hartley1928,Shannon1948,Cherry1963}.

In this paper, we focus on the implications of this for the \textit{average} length of a component in such a system.  In the proteome, this would be the average length of a protein in amino acids and in a computer function, this is the average length in the discrete tokens of the programming language in which it is written \cite{HattonWarr2017}.

\begin{figure}[ht!]
\centering
\includegraphics[width=0.5\textwidth]{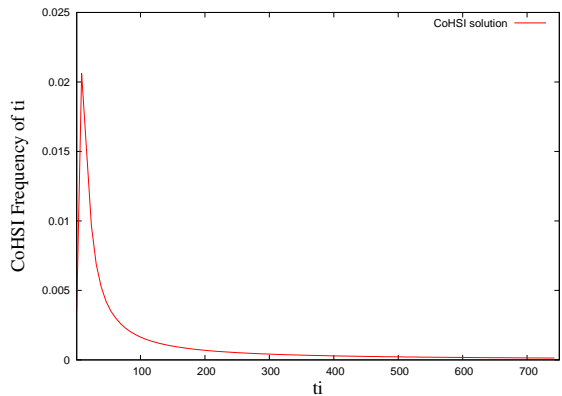}
\caption{\label{fig:mdata}Illustrating a typical solution of the CoHSI equation described in \cite{HattonWarr2017}.}
\end{figure}

A typical solution of (\ref{eq:minif}) is shown as Figure \ref{fig:mdata}.

\begin{figure}[ht!]
\centering
\includegraphics[width=0.5\textwidth]{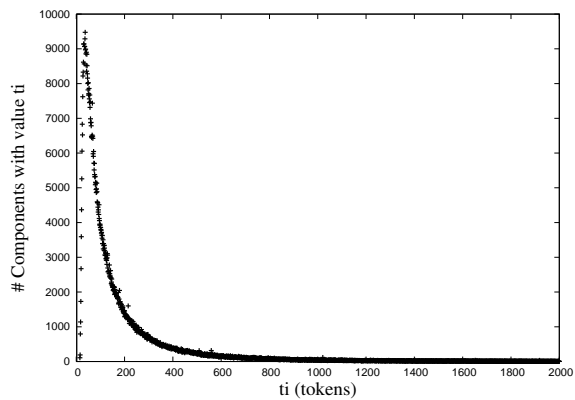}
\caption{\label{fig:cdata}Illustrating the length distribution of the functions measured in programming language symbols for 80 million lines of open source C.}
\end{figure}

\begin{figure}[ht!]
\centering
\includegraphics[width=0.5\textwidth]{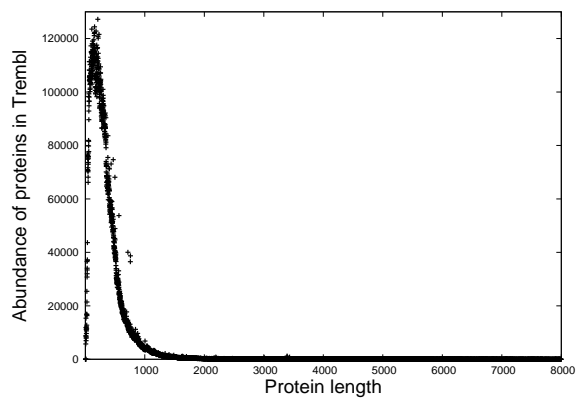}
\caption{\label{fig:tdata}Illustrating the length distribution of the proteins measured in amino acids in TrEMBL version 15-07.}
\end{figure}

This can be compared with actual length distributions such as those of collections of computer software and also the collection of all known sequences (the pan-proteome).  Figure \ref{fig:cdata} shows a plot of the frequency of occurrence of different lengths of functions measured in programming language tokens for 80 million lines of open source software written in the programming language ISO C and Figure \ref{fig:tdata} shows the length of proteins measured in amino acids for the known proteome as defined by TrEMBL version 15-07 \cite{Trembl2018}.  As well as their striking visual similarity, they are mathematically very similar; both are characterized by a sharp unimodal peak with almost linear slopes, asymptoting to an extraordinarily accurate power-law \cite{HattonWarr2015,HattonWarr2017}.
%
%

The fact that such a model exists and accurately models measured length distributions in such disparate systems means that we can throw light on some interesting questions relating to the average length of a protein (including what average even means for this distribution) in light of the continuous changes in protein sequences that the processes of evolution are acknowledged to generate.  For the remainder of this paper, we will therefore focus on the proteins.

\section{Conservation of average protein length}
Researchers have previously noted the highly conserved nature of the average protein length in different aggregations.  For example, \cite{Wang2005} showed that there was a general tendency for protein lengths to be conserved across the eukaryotic domain, whilst noting that protein orthologs were of different average length across domains.  \cite{XuJune2006} noted that in their collections, the average lengths of genes are highly conserved in both prokaryotes and eukaryotes although the average length in the two domains are rather different. Given the fact that CoHSI is a constraint that acts on the global properties (including the lengths) of proteins in all organisms, the variations in average length introduced above are of obvious interest.  By utilizing the properties of the underlying CoHSI length distribution as described in \cite{HattonWarr2018a}, we will examine these observations; in particular we will introduce the notion of departure from the equilibrium or most likely state of average protein length as a measure of \textit{evolutionary divergence}. Initially we will explore the concept of the "average length" of proteins  in different aggregations  of protein sequences taken from a recent version of the best-annotated database, SwissProt (version 18-02, \url{https://uniprot.org} - we have used various versions of SwissProt to demonstrate the robustness of our results).  

\subsection{Understanding average protein length}
The distribution of Figure \ref{fig:mdata} as exemplified in the real systems of Figures \ref{fig:cdata} and \ref{fig:tdata} has interesting properties.  Although it is the solution of a differential equation, for components which are large compared with their unique alphabet of tokens (e.g. amino acids, programming language tokens ...), it asymptotes quickly to a power-law, (typically from $t_{i} > 50$ although it depends on the size of the unique alphabet of tokens used by the system under consideration).  In the real data of Figures \ref{fig:cdata} and \ref{fig:tdata} the power-law is astonishingly accurately produced, (giving adjusted $R^{2}$ values in R of > 0.99 \cite{HattonWarr2017}).  For smaller components, the distribution is characterized by a sharp unimodal peak with almost linear slopes.

The presence of the power-law means that the distribution is long-tailed and along with the sharp unimodal peak at small $t_{i}$, leads to a distribution which is palpably right-skewed.  This significantly complicates the meaning of the word \textit{average}.  Statisticians use the words \textit{location} and \textit{spread} to study distributions.  Location is in a very general sense what laymen call the average.  It is a description of "where" a distribution appears to be concentrated.  Spread refers to how much the distribution spreads around this location.  Hence when we use the word average, we have in mind the most likely value if presented with a member of the population we are considering.  All readers will be familiar with the normal distribution or bell curve, which is a symmetric unimodal distribution and therefore its most common value is in the middle at the peak.  For such a distribution, the standard measures of average are the mean, median and mode.  For a symmetric, unimodal distribution the values of these three measures are coincident, thus fitting in comfortably with our notion of average.  However, for a skewed distribution such as shown in Figure \ref{fig:mdata}, the mean, median and mode values are not coincident.  In the case of the proteome, this important distinction has not escaped some previous researchers \cite{Zhang2000} but it is a point worth making when trying to understand what exactly is highly conserved.

By way of example and to show that mean, median and mode can differ substantially, if we compute each of these for the length distributions of the domains of life and viruses in TrEMBL release 15-07 of Figure \ref{fig:tdata} (data derived from \url{https://uniprot.org}), we get the values shown as Table \ref{tab:average}.
%
%

\begin{table}[h!]
\centering
\caption{Measures of average length of proteins in the domains of life and viruses}
\begin{tabular}{@{\vrule height 10.5pt depth4pt  width0pt}cccc}
\hline
Domain of Life & Mean & Median & Mode \\
\hline
Archaea & 287 & 246 & 130 \\
Bacteria & 312 & 272 & 156 \\
Eukaryota & 435 & 350 & 379 \\
\hline
Viruses & 451 & 289 & 252 \\
\hline
\end{tabular}
\label{tab:average}
\end{table}

The effects of the skew are obvious and as can be seen, the mean is probably the least satisfactory way of measuring the average.  The median (in terms of the spread of its values in Table \ref{tab:average}) is less sensitive to the skew than are the mean and the mode, and is similar across all domains of life (even for viruses) although with the highest value for the Eukaryota.  While there are other statistically appropriate ways of dealing with this skew under the general banner of robust methods such as the trimmed mean, (c.f. \cite{Tukey1977}), whereby we remove some percentage of the population at both the low and the high end of the distribution, we will not consider these further here.  The data shown in Table 1 give clear support for the observations previously made by others that conventional measures of the "average" are indeed well conserved in the proteins; however, here we can ascribe this property to the  conservation principle (CoHSI) that constrains the nature of the underlying length distributions.

As noted above, Table \ref{tab:average} has been interpreted as suggesting a significant difference in the mean length found in prokaryota and eukaryota in \cite{XuJune2006}.  We can see immediately here, that whatever conventional measure of average is used, eukaryota do in fact shows a higher average, with variations in the median being rather less than those of the mean, as would be expected from such a skewed distribution.  \textit{In other words, this observation is robust with respect to the measures of average used.}

This allows us to use a useful graphical property of the mean.
%
%
%
%
%
%
%
%

\subsection{Exploring average protein length in different aggregations}
In this section, we will explore how protein lengths vary across aggregations of protein data taken at different levels in the taxonomic hierarchy.  Following the sage advice of Tukey \cite{Tukey1977}, we will begin by doing this visually.  One very useful device for this is to plot the total number of proteins $N$ against the total concatenated length $L$ of those proteins (in amino acids) for each species in an aggregation.  If we use the mean $m$ as a measure of the average length of proteins in a species, then $L = m.N$, and so plotting $L$ against $N$ gives a straight line of gradient $m$ if the average length is the same for all species in the plot.

\begin{figure}[ht!]
\centering
\includegraphics[width=0.5\textwidth]{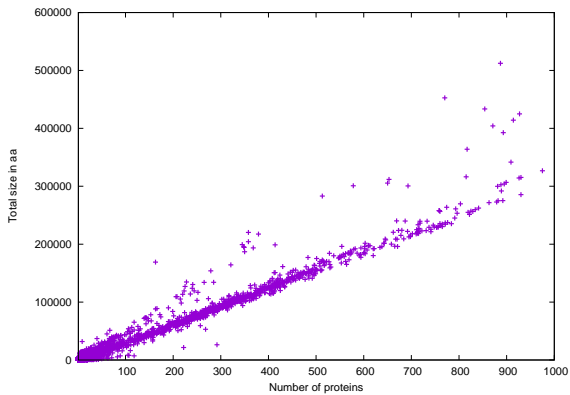}
\caption{\label{fig:all}A plot of the total length of all proteins in a species against the number of proteins in a species for the whole of the SwissProt dataset version 18-02.}
\end{figure}

Figure \ref{fig:all} illustrates this for the whole of the SwissProt annotated subset of TrEMBL, release v18-02 for all species with up to 1,000 proteins.   Each point corresponds to a species and represents the total length of all the proteins in that species plotted against the total number of proteins for that species.  Similar data are shown for SwissProt release 13-11 in \cite{HattonWarr2015}.  As can be seen, the resulting distribution is highly linear indicating that the mean length of proteins is indeed strongly conserved across the entire SwissProt release.

\subsubsection{The domains of life}
\begin{figure}[ht!]
\centering
\includegraphics[width=0.5\textwidth]{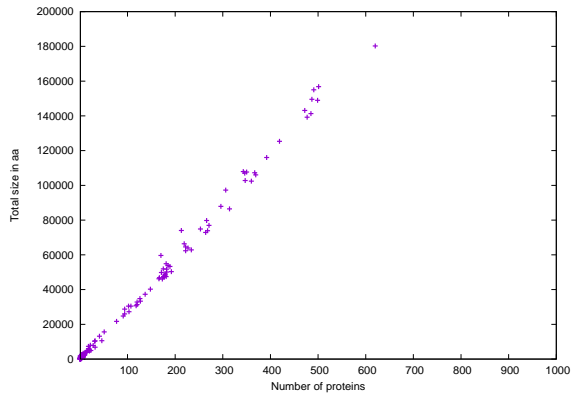}
\caption{\label{fig:archaea}A plot of the total length of all proteins in a species against the number of proteins in a species for the archaea of the SwissProt dataset version 18-02.}
\end{figure}

\begin{figure}[ht!]
\centering
\includegraphics[width=0.5\textwidth]{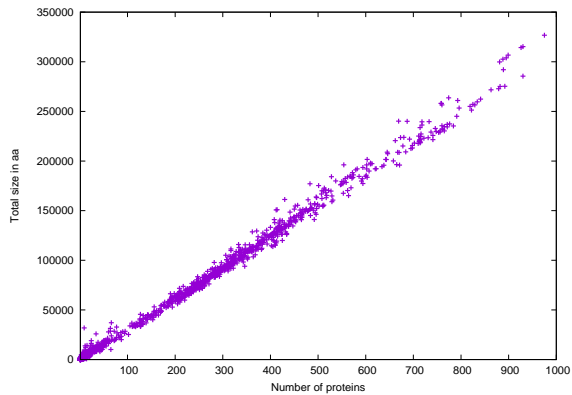}
\caption{\label{fig:bacteria}A plot of the total length of all proteins in a species against the number of proteins in a species for the bacteria of the SwissProt dataset version 18-02.}
\end{figure}

Figures \ref{fig:archaea} and \ref{fig:bacteria} show the equivalent plots for the archaea and the bacteria of the SwissProt release v18-02 respectively.  These are subsets of Figure \ref{fig:all}.
%
%

\begin{figure}[ht!]
\centering
\includegraphics[width=0.5\textwidth]{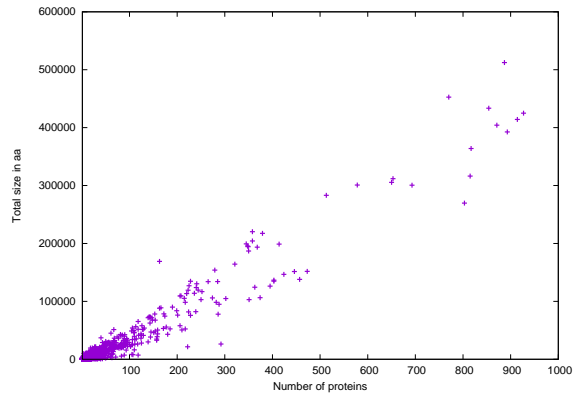}
\caption{\label{fig:eukaryota}A plot of the total length of all proteins in a species against the number of proteins in a species for the eukaryota of the SwissProt dataset version 18-02.}
\end{figure}

In contrast, Figure \ref{fig:eukaryota} shows the equivalent plot for the eukaryota.  Although the overall behaviour is strongly linear as predicted, there appear interesting departures, e.g. the broadening tending to bifurcation of the plot, which do not seem to be present in either archaea or bacteria.

\subsubsection{Viruses}
\begin{figure}[ht!]
\centering
\includegraphics[width=0.5\textwidth]{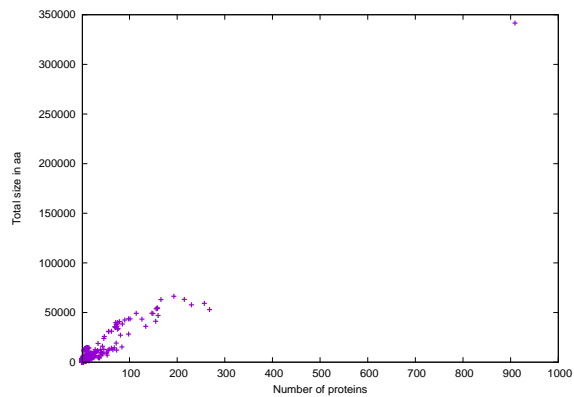}
\caption{\label{fig:viruses}A plot of the total length of all proteins in a species against the number of proteins in a species for the viruses of the SwissProt dataset version 18-02.}
\end{figure}

For completeness, we also include the viruses as Figure \ref{fig:viruses}. We will however say little more about these data , as the status of viruses as non-living agents that infect cells (and co-opt their biochemical and cellular machinery) across the whole spectrum of the domains of life introduces additional complications and questions that we do not address. In particular, the viruses do not represent a natural aggregation; rather they are a heterogeneous collection that might best be considered as a component of  the domain of life that they specifically infect.
%

Note that we do not expect the power-law slopes of the tails of these different aggregations to be the same.  In terms of our model, they correspond to different values of $\alpha, \beta$ but our theory has nothing to say about their actual values in any system as they are fundamentally undetermined in the Statistical Mechanics methodology used.

We give an R lm() linearity analysis for each of the aggregations in Figs \ref{fig:all} - \ref{fig:viruses} as Table \ref{tab:average2}.  Note that these are somewhat different than those quoted in Table \ref{tab:average} which used v.15-07 of the SwissProt annotated subset.

\begin{table}[h!]
\centering
\caption{Measures of mean length of proteins in the domains of life and viruses taken from the SwissProt annotated subset of TrEMBL v.18-02.}
\begin{tabular}{@{\vrule height 10.5pt depth4pt  width0pt}ccccc}
\hline
Aggregation & Slope & Adjusted $R^{2}$ & Std. Error & \# species \\
\hline
All & 491.5 & 0.968 & 0.764 & 13,590 \\
Archaea & 283.1 & 0.993 & 1.626 & 212 \\
Bacteria & 312.9 & 0.997 & 0.336 & 3,000 \\
Eukaryota & 520.0 & 0.989 & 0.629 & 7,751 \\
\hline
Viruses & 344.4 & 0.926 & 1.903 & 2,627 \\
\hline
\end{tabular}
\label{tab:average2}
\end{table}

The adjusted $R^{2}$ values of Table \ref{tab:average2} confirm our visual analysis with prokaryotic domains being very close to the equilibrium value (which corresponds to 1.0).  Viruses show the biggest departure.

\section{Evolutionary divergence and fine structure in the eukaryota}
We will now attempt to interpret the visual patterns of Figures \ref{fig:all}, \ref{fig:archaea}, \ref{fig:bacteria}, \ref{fig:eukaryota} and, although the viruses are not considered a domain of life (but still manifestly a qualifying discrete system), Figure \ref{fig:viruses}, in terms of our theory.

As described at length in \cite{HatTSE14,HattonWarr2017}, our theory is scale-independent and token-agnostic.  As a result, we argue with very substantial measurement support \cite{HattonWarr2015}, that discrete systems \textit{inter alia} share common properties associated with the distribution of the lengths of their components.  These properties should therefore by definition be decoupled from the particular nature of the system.  In other words, in the case of the proteome, it is unnecessary to postulate that particular length distributions need to have been generated by evolution, which is brought about by a variety of local processes, only some of which rely on the meaning and therefore function of tokens (amino acids). However, we note that evolutionary processes can perturb the equilibrium distribution constrained by CoHSI - a measure of the degree of this perturbation is the divergence from linearity in the plots of Figs 4-8, as expressed for example in the adjusted $R^{2}$ values of Table \ref{tab:average2}.

It is helpful to elaborate on the concept of evolutionary divergence as we use it here, in order to emphasize that we  do not see evolution as being driven solely by the pressure of natural selection. The modern synthesis (or neo-Darwinian theory) regards evolution (simply, the change of life forms with time) as a process with a primary mechanism of natural selection. Naturally arising or pre-existing variants in a population are acted on by natural selection, such that the best-adapted organisms (through the survival of the fittest) pass their conditionally superior genes to the next generation. However, genetic variation in a population can also be the basis for evolution by mechanisms distinct from natural selection. These mechanisms operate in one case by the fixation in a population of particular genetic variants through what are essentially stochastic processes. Because the variants fixed in this random manner are postulated to be without substantial benefits or handicaps in terms of natural selection, this theory is termed \textit{genetic drift}, or the neutral theory of evolution. Neutral evolution may even explain the majority of genetic variation that is seen at the molecular level, \cite{Kimura1989,Stoltzfus2012}.

A second mechanism that acts to fix particular genetic variants in a population  and that is not driven by natural selection is termed \textit{molecular drive}. In this theory \cite{Dover1982}, intrinsic mechanistic, molecular and biochemical biases in cellular functions lead to an outcome where particular genetic variants become fixed in a population without the process being driven by natural selection. Processes that are susceptible to these biases include gene conversion (which can show directional preferences), crossing-over that occurs preferentially at certain chromosomal sites bearing specific allelic variants, and chromosomal rearrangements such as transpositions, which can show preferences for certain sites.   Note that we cannot eliminate the role of any particular mechanism of evolution in the observed patterns of protein length - such a proposal would not be falsifiable.  All we can say is that it is not \textit{necessary} to invoke specific evolutionary mechanisms to explain what is observed, in an analogous way to the role of the ether in early 20th century discussions of relativity.

Returning to our discussion of the aggregations of the previous section, there is a clear visual difference between the length distributions of the prokaryotic domains of life (archaea and bacteria) when compared with the eukaryotic domain of life.  The former are more or less exactly what we expect of a system which closely adheres to the CoHSI principle.  In \cite{HattonWarr2018a}, we explored the general properties of (\ref{eq:minif}), in particular, the nature of the solutions as the Lagrangian undetermined parameters $\alpha, \beta$ were varied and the effects of these variations on three measures of the average, i.e. the mean, the median and the mode.  We showed that all three measures were robust to variations in $\alpha, \beta$ consistent within ranges of values observed in real data and specifically how they were related to the power-law slope of the tail of the length distribution.

\textit{In other words, any system obeying (\ref{eq:minif}) would be expected to exhibit a strongly conserved average component length across different aggregations, howsoever measured, simply because variations in the disposable parameters $\alpha, \beta$ for those different aggregations have relatively little effect on the mean, median and mode.  Indeed, strongly conserved average component length is a property of CoHSI systems.}

How then should we interpret the data for eukaryota  shown in Figure \ref{fig:eukaryota} in terms of the apparent bifurcation of their mean protein length?  Using an analogy from physical systems, we can think of the solution of (\ref{eq:minif}) as the \textbf{equilibrium state} and any departures from it as due, in the case of the proteome, to \textbf{evolutionary pressure}.  We could then interpret the notable visual features of Figure \ref{fig:eukaryota} either as characteristic of a \textit{less efficient domain of life} where return to the equilibrium state (exactly conserved average length) is more sluggish, or as a domain of life where evolutionary processes are exerting a greater pressure. At this point we merely note these two possible explanations, and prefer not to engage in untestable speculation about defining in quantitative terms what \textit{efficiency} could mean in terms of the cellular and molecular processes of the different domains of life, or the relative rigors of selective pressure experienced by archaea, bacteria and eukaryota.  

Returning then to the prokaryotic domains, we note that the archaea and bacteria both closely adhere to the expected behavior for a discrete system, that of close adherence to the conservation of average protein length.  In other words, both prokaryotic domains of life appear close to the equilibrium state with little evidence of departures other than what appear to be minor random fluctuations.  While this outcome would be consistent with speculation that prokaryotes
are both 1) subject to strong selective pressure and 2) possess highly efficient mechanisms of response to such pressure, a consequence of the approach that we take here is that there is no \textit{prima facie} reason to look for evolutionary implications when the only departures from the equilibrium state constrained by CoHSI appear both minor and random.

When looking at the eukaryota of Figure \ref{fig:eukaryota} however, our visual impression is that the departures are \textit{not} random and there appears evidence of fine structure in the form of bifurcated embedded regions of linearity with different gradients, almost as if the eukaryota could be further sub-divided.  There is also a hint of this in the full dataset itself as shown in Figure \ref{fig:all} with a zone of steeper linearity evident with a gradient corresponding to an average protein length of around 440.   Now we recall \cite{HattonWarr2017}, that although the CoHSI principle is overwhelmingly likely to lead to a pdf which is the solution of (\ref{eq:minif}), \textit{it is not a straitjacket.}  We expect in a CoHSI system that average component length will be strongly conserved across different aggregations, so it is of considerable interest to investigate apparently systematic departures from this as appear to occur in the eukaryota.   \textit{We therefore propose that systematic departures from CoHSI both identify and provide a measure of evolutionary divergence, which we now test.}

\subsection{Incompleteness}
Before we begin to explore the dataset itself, it is important to discuss what we mean by systematic departures in these large datasets.  All experimental datasets include various kinds of noise, such as pseudo-random noise perhaps caused by data which has not yet been curated properly, but also systematic noise caused for example by researchers choosing to study only the small proteins of a species.  We can do some simple analyses to explore these points.

For example, if we took a minimum qualifying number of proteins even as low as 500 in order to measure how well researchers have covered a particular species, then more than 98\% of the species appearing in the better curated SwissProt dataset would \textbf{not} qualify.  Even if the very much larger TrEMBL v 18-02 is used (larger by about a factor of 100x), almost 85\% of the species would not qualify either.

\begin{figure}[ht!]
\centering
\includegraphics[width=0.5\textwidth]{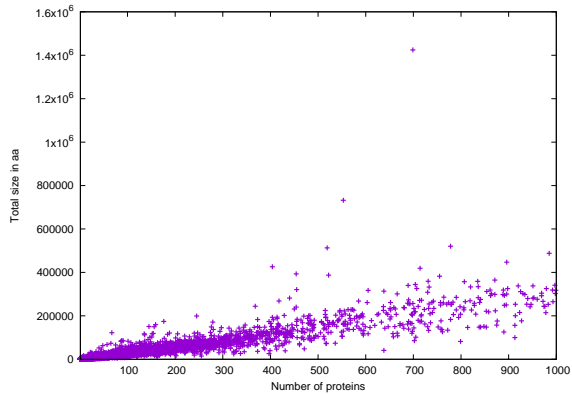}
\caption{\label{fig:tremblall}A plot of the total length of all proteins in a species against the number of proteins in a species for all species in the Trembl dataset version 18-02.}
\end{figure}

Figure \ref{fig:tremblall} is a plot of all species in the less well curated superset TrEMBL v 18-02.  It can be directly compared with the better curated SwissProt 18-02 subset Figure \ref{fig:all} as the x-axis scale is the same in both, (although not the y-axis).

Comparing Figures \ref{fig:all} and \ref{fig:tremblall} gives some intriguing insights into this question.  First of all it is clear that the bifurcation observed in Figure \ref{fig:all} is completely obscured by the additional noise in the full TrEMBL distribution Figure \ref{fig:tremblall}. If we interpret Figure \ref{fig:all} as being indicative of genuine biological signal there are hints of systematic behaviour but as we will see, we have to reduce our target qualifying number of proteins dramatically to only 40 before species emerge, but do they do so consistently ?

\subsection{Visual tools for fine structure analysis}
CoHSI is a theory about tokens in discrete systems and the direct implications for how length distributions of proteins, in the case of the proteome, will behave.  In our analysis of fine structure, we will therefore use graphical properties of the length distributions themselves, notably box and whiskers diagrams which show the outliers, the quartiles and the median, and give more insight into the skewed nature of the distribution.

We observe in our analyses what appear to be two patterns of local linearity in Figure \ref{fig:all} associated with the eukaryota, as seen more clearly  in Figure \ref{fig:eukaryota}, where the subset of data for the eukaryota are plotted separately. Two populations emerge, clearly distinguishable by their average length.

\begin{enumerate}
\item High average protein length eukaryota $440 \pm 40$ aa.
\item Population average protein length eukaryota $300 \pm 30$ aa.
\end{enumerate}

The large majority of points in Figure \ref{fig:eukaryota}  (each corresponding to a species), lie on or near what we interpret as the equilibrium state for the system and which we have named the \textit{Population average protein length}, in this case using the mean.  This is where the vast majority of species are located.  However, the second population of interest also shows marked linearity  and is termed the \textit{High average protein length}; this we will focus on.  It is visible as an approximately linear band with systematically higher gradient than the population average.  Since it appears to have a consistent quality in spite of the relatively low number of qualifying proteins we have been forced to use by incompleteness, we would expect that a biological signal would correspond to some systematic set of species occupying this zone of higher average length.  On the other hand, if the zone of linearity appears to have no consistency with respect to species, we would have much less confidence in it as a distinguishable population.

\subsubsection{High average protein length eukaryota}
\begin{figure}[ht!]
\centering
\includegraphics[width=0.8\textwidth]{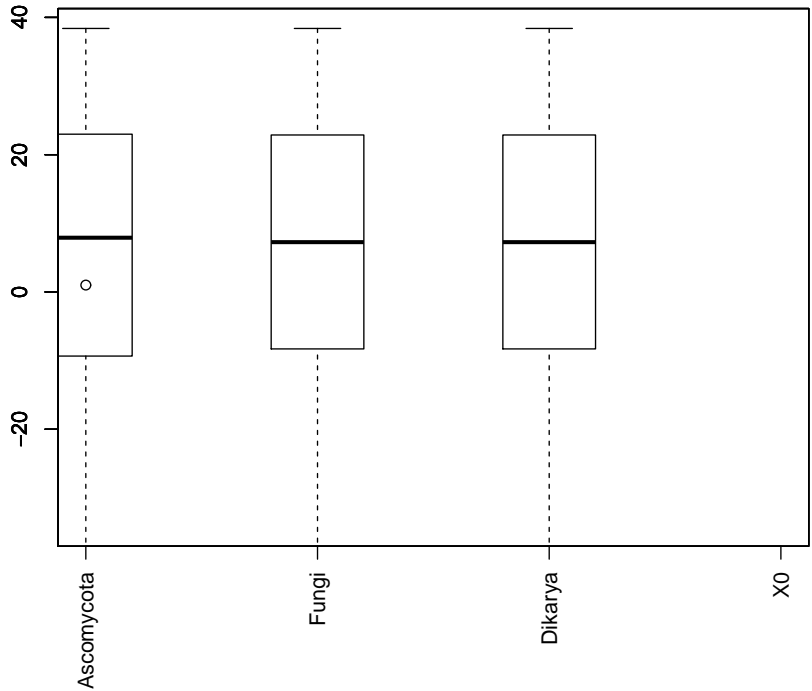}
\caption{\label{fig:higheukaryota_40} A box and whiskers diagram of all eukaryota with an average protein length of $440 \pm 40$ aa for a minimum qualifying number of proteins of 40.}
\end{figure}

Figure \ref{fig:higheukaryota_40} shows a box and whiskers diagram for the range of eukaryota with the higher  average protein length of $440 \pm 40$ with the minimum number of qualifying proteins set at 40.  We see that the species occupying this zone of linearity are indeed consistent and are members of the kingdom Fungi and specifically the subkingdom dikarya and the phylum ascomycota.  In spite of the heavy qualification resulting from the incompleteness, this is sufficiently promising that we will continue dropping the minimum number of qualifying proteins to see at what point species from a different kingdom intrude, clouding our picture.

\begin{figure}[ht!]
\centering
\includegraphics[width=0.8\textwidth]{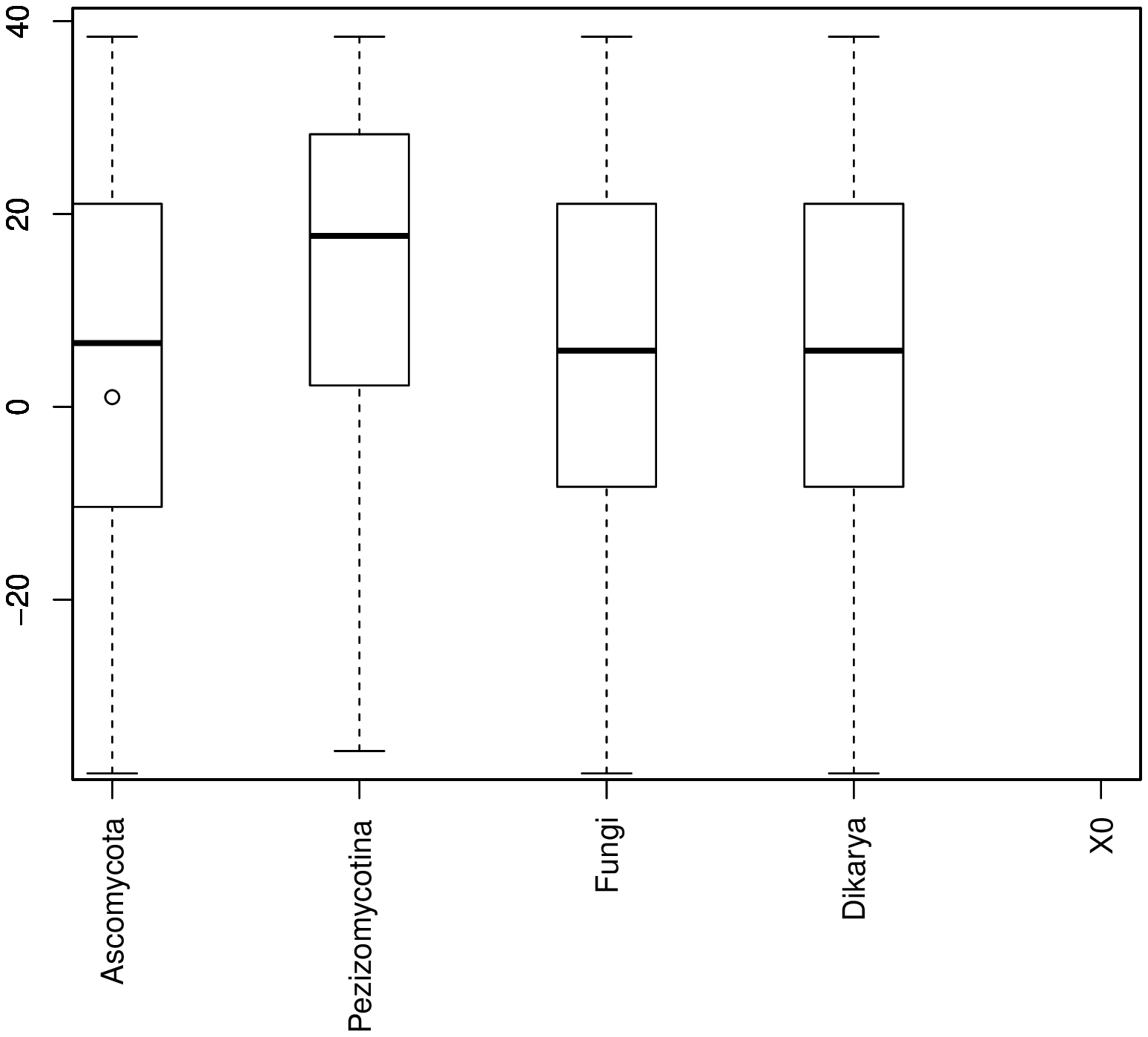}
\caption{\label{fig:higheukaryota_30} A box and whiskers diagram of all eukaryota with an average protein length of $440 \pm 40$ aa for a minimum qualifying number of proteins of 30.}
\end{figure}

Figure \ref{fig:higheukaryota_30} is the same data but with the addition of species with a qualifying number of proteins set to 30 or greater.  This time a new subphylum pezizomycotina  appears - notable that this is the largest subphylum of the ascomycota fungi (http://tolweb.org/Pezizomycotina/29296).

%
%

\begin{figure}[ht!]
\centering
\includegraphics[width=0.8\textwidth]{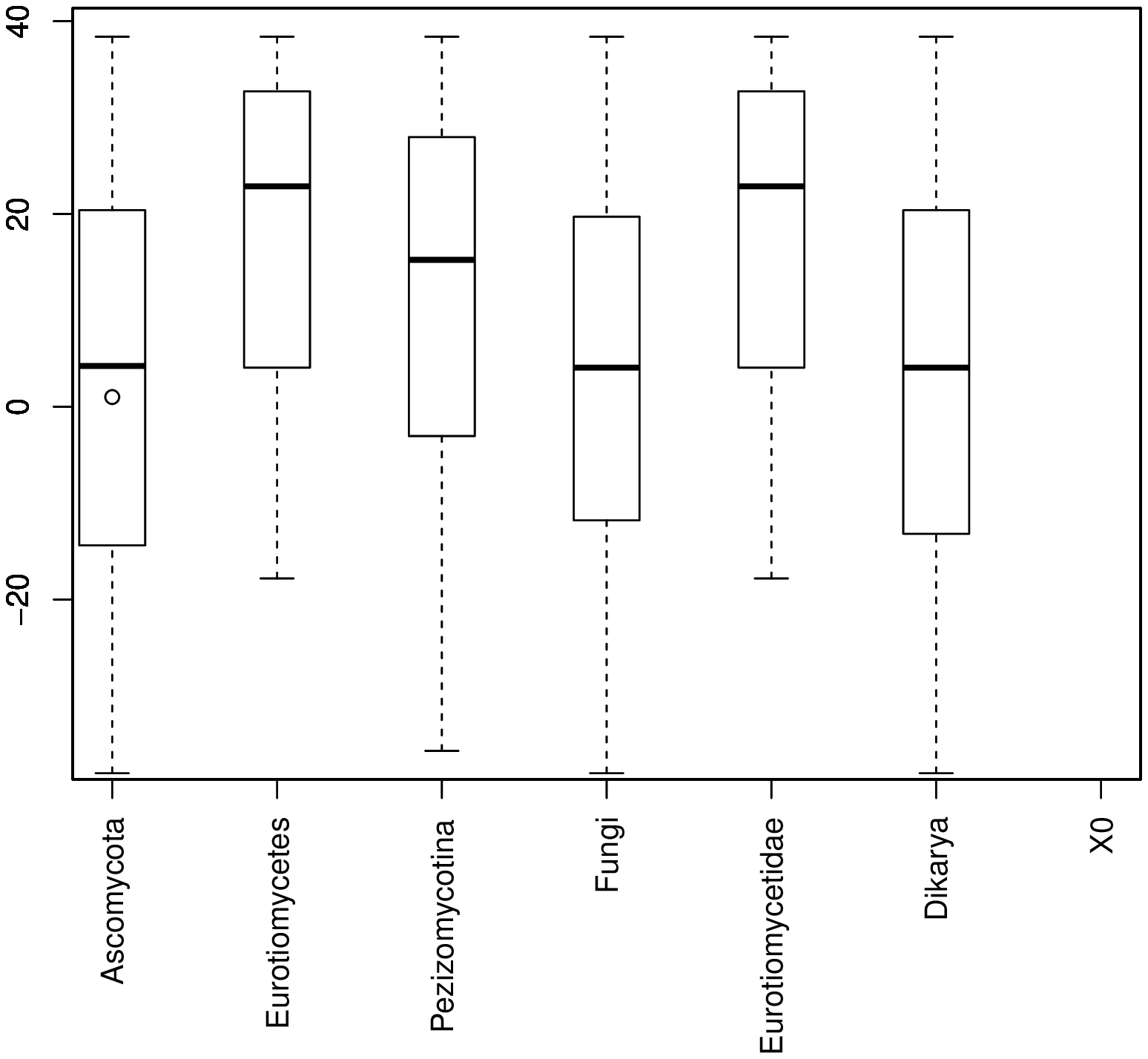}
\caption{\label{fig:higheukaryota_20} A box and whiskers diagram of all eukaryota with an average protein length of $440 \pm 40$ aa for a minimum qualifying number of proteins of 20.}
\end{figure}

Once again dropping the minimum number of qualifying number of proteins to 20, Figure \ref{fig:higheukaryota_20} results.  Again only members of the kingdom fungi are introduced, this time eurotiomycetes, a class of pezizomycotina and eurotiomycetidae, a subclass of the eurotiomycetes.
%
%

Only when we reduce below this already low level of qualifying proteins (data not shown) are additional species from taxa other than the ascomycota (such as the Metazoa) identified.

Our results suggest that deviation from the equilibrium protein length distribution  constrained by CoHSI can potentially identify evolutionary divergence, as is the case documented here for the ascomycete fungi. The fact that we could reduce the minimum number of qualifying proteins for a particular taxon down to only 20 whilst preserving the result shows a level of consistency which is certainly promising and might be indicative of a biological signal. The nature of  any underlying evolutionary mechanism and possible significance for the ecology and evolution of the ascomycota could be a fruitful area of investigation.
%
%

\subsection{A preliminary speculation on the cause of fine structure}
We have identified above that it seems possible to identify particular species associated with differing average protein length within their domain aggregation.  What might be the cause of this?

At this stage, we will do no more than identify one further interesting property in information based discrete systems which is \textit{not} shared with the classical Boltzmann systems in the Kinetic Theory of Gases.  Recall from \cite{HattonWarr2017} that in the review of classical Boltzmann systems, the parameter $\alpha$ controls the total size $T$ of the system in tokens whilst the parameter $\beta$ controls the exponential shape of the energy distribution.  However in that theory, $\alpha, \beta$ are tightly coupled.  Once $\beta$ is chosen, $\alpha$ is set automatically as it simply normalises the distribution, including subsets of $T$.  \textit{In other words, subsets have the same shape if they have the same $\beta$.}  (Note that in classical Boltzmann systems, $\beta = 1/(k_{B}\Theta)$ where $k_{B}, \Theta$ are Boltzmann's constant and the temperature respectively.)

In information-based systems however, $\alpha, \beta$ \textbf{decouple}, \cite{HattonWarr2018a}.  The parameter $\alpha$ still controls the system size $T$ and the parameter $\beta$ controls how the information is distributed, but their functionality now overlaps.  \textit{$\beta$ is still determined asymptotically by the shape of the distribution (the distribution of the unique alphabet of amino acids in this case), but this does not automatically determine $\alpha$.  Instead there are a range of distribution shapes for the smaller components corresponding to different values of $\alpha$ for the same value of $\beta$, (Figure 6 of \cite{HattonWarr2018a}).}  This corresponds to varying the system size for the same total information content - the different distributions naturally lead to different values of the average for subsets even with the same $\beta$.  There is no analog for this in classical energy conserving systems.

We can illustrate this by considering only the Bacteria domain of life in version 18-02 of the SwissProt dataset.  We have already seen in Figure \ref{fig:bacteria} that this dataset appears to adhere very closely to the highly conserved average protein length described earlier as an equilibrium state, with no obvious fine structure, unlike that observed in the eukaryota.  If this were a classical system, subsets of this dataset would have the same statistical properties and we would expect them to have the same albeit somewhat noisier estimated average length.  However, this is not the case with CoHSI systems which have a richer set of subtle behaviours.  We can see this by extracting subsets based on their \textit{unique amino acid alphabet count}, one of the key parameters of discrete systems to emerge in \cite{HattonWarr2017}.  Now for fixed $\beta$, smaller subsets will correspond to smaller $\alpha$ since this parameter controls the size but as we showed in \cite{HattonWarr2018a} the CoHSI equation then implies a shorter average protein length.

\begin{figure}[ht!]
\centering
\includegraphics[width=0.5\textwidth]{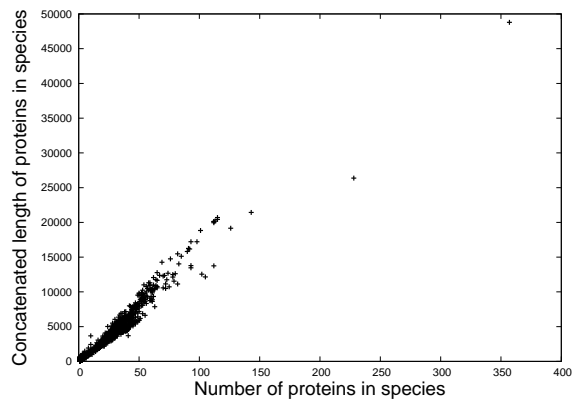}
\caption{\label{fig:bacteria18}A plot of the total length of all proteins in a species against the number of proteins in a species for all bacteria of the SwissProt dataset version 18-02 with exactly 18 unique amino acids.}
\end{figure}

\begin{figure}[ht!]
\centering
\includegraphics[width=0.5\textwidth]{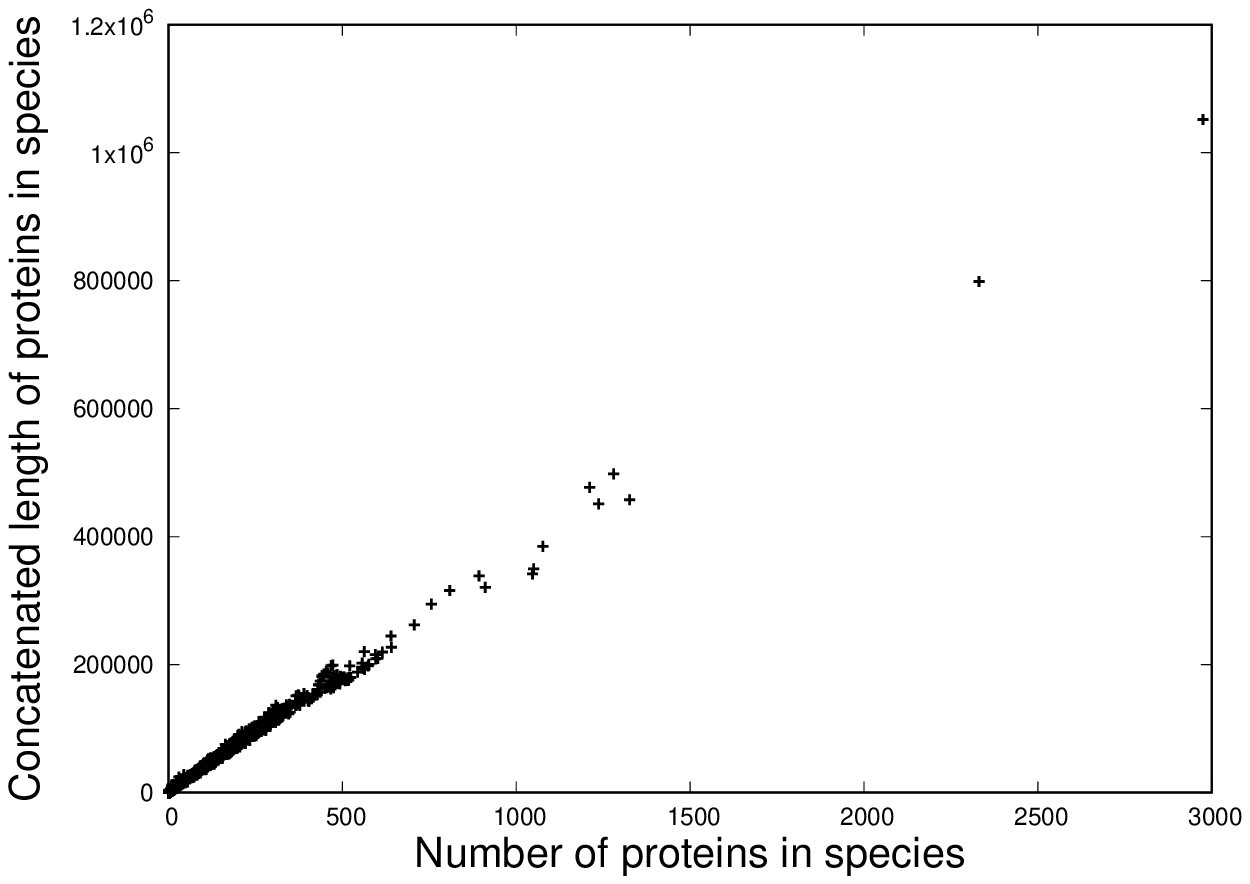}
\caption{\label{fig:bacteria20}A plot of the total length of all proteins in a species against the number of proteins in a species for the bacteria of the SwissProt dataset version 18-02 with exactly 20 unique amino acids.}
\end{figure}

We can clearly see this prediction fulfilled qualitatively in Figures \ref{fig:bacteria18} and \ref{fig:bacteria20}, (they are typical of analyses we conducted for 12-24 unique amino acids in increments of 1).  According to an R lm() analysis, the 18 unique amino acid count proteins have an average length of 152.0, adjusted $R^{2}$ of 0.955 on 1241 species whilst the 20 unique amino acid count proteins have an average length of 365.5, adjusted $R^{2}$ of 0.995 on 2544 species.  This is a good point to re-emphasize the token-agnostic nature of CoHSI.  It simply does not matter which amino acids are actually used since they have no intrinsic meaning.  The only thing that matters in the information measure is the \textit{unique} amino acid count \cite{HattonWarr2015,HattonWarr2017}.

We will defer any further discussion of average protein length and unique alphabet size as it properly belongs in a full discussion of the complex topic of protein alphabets and PTM (Post-Translational Modification) which we will consider in detail in a later paper in this series.  We do however consider Figures \ref{fig:bacteria18} and \ref{fig:bacteria20} as further supporting evidence of the CoHSI theory.

\section{Conclusions}
In this paper, we have discussed various measures of average length in CoHSI systems showing that observed properties of the proteome notably the strong conservation of average lengths of proteins should be expected whatever standard measure of average length we use, be it mean, median or mode as indicated in \cite{HattonWarr2018a}. 

Choosing the mean and using a simple graphical property, we then demonstrated that this leads to a consistent visual method of identifying related taxa in an average length plot.  The archaea, bacteria and eukaryota each show distinct average protein length distributions, and a closer examination of the eukaryota revealed that fungi showed a longer then average protein length. Examining the fungi in more detail suggested that this property of the fungi might be ascribed specifically to the phylum ascomycota, although we caution that the current coverage of protein sequencing for many species is quite minimal.  This situation improves all the time as sequencing and annotation efforts continue, and the availability of more comprehensive and reliable datasets for proteins will improve the confidence in results obtained through the methods outlined here.
%
%

We also introduced the idea of evolutionary divergence (or pressure) acting against an equilibrium state defined by CoHSI.  This was particularly well exemplified in comparing the rather noisy full length distributions of all species in TrEMBL v 18-02 with those in the better curated and much smaller subset, SwissProt v 18-02.  This latter dataset allowed us to identify what we consider as biologically-related fine structure in the eukaryota in the form of banded zones of linearity in the average protein length of species.  This was not visible in archaea or bacteria, which we hypothesize are much closer to the CoHSI- constrained equilibrium.

Finally, we speculated that there may be some relationship with unique amino acid alphabet as predicted by \cite{HattonWarr2017} and we were able to demonstrate the existence of a subtle relationship directly predicted by CoHSI on the Bacteria domain of life in version 18-02 of SwissProt whereby different sized subsets based on unique amino acid count of this domain have different average lengths.  We will pursue the nature of this relationship and the post-translational modification of proteins in a later paper in this series.

\clearpage

\bibliographystyle{alpha}
\bibliography{bibliography}

\end{document}